\documentclass[twocolumn,english,superscriptaddress,floatfix]{revtex4}
\usepackage[T1]{fontenc}
\usepackage[latin9]{inputenc}
\usepackage{color}
\usepackage{bm}
\usepackage{babel}
\usepackage{amsmath}
\usepackage{amssymb}
\usepackage{esint}
\usepackage{graphicx}
\usepackage[unicode=true,
 bookmarks=false,
 breaklinks=true,pdfborder={0 0 1},backref=section,colorlinks=true]
 {hyperref}
\hypersetup{
 pdfcreator={},pdfproducer={LaTeX with hyperref},linkcolor=blue,anchorcolor=blue,citecolor=blue,filecolor=red,menucolor=red,pagecolor=red,urlcolor=blue,pdfstartview=FitV,pdfhighlight=/I,pdfpagelayout=OneColumn,hypertexnames=true}

\makeatletter

\newcommand*\LyXThinSpace{\,\hspace{0pt}}

\@ifundefined{textcolor}{}
{%
 \definecolor{BLACK}{gray}{0}
 \definecolor{WHITE}{gray}{1}
 \definecolor{RED}{rgb}{1,0,0}
 \definecolor{GREEN}{rgb}{0,1,0}
 \definecolor{BLUE}{rgb}{0,0,1}
 \definecolor{CYAN}{cmyk}{1,0,0,0}
 \definecolor{MAGENTA}{cmyk}{0,1,0,0}
 \definecolor{YELLOW}{cmyk}{0,0,1,0}
}

\usepackage{babel}

 \@ifundefined{textcolor}{}{%
  \definecolor{BLACK}{gray}{0}
  \definecolor{WHITE}{gray}{1}
  \definecolor{RED}{rgb}{1,0,0}
  \definecolor{GREEN}{rgb}{0,1,0}
  \definecolor{BLUE}{rgb}{0,0,1}
  \definecolor{CYAN}{cmyk}{1,0,0,0}
  \definecolor{MAGENTA}{cmyk}{0,1,0,0}
  \definecolor{YELLOW}{cmyk}{0,0,1,0}
 }

 \@ifundefined{textcolor}{}{
  \definecolor{BLACK}{gray}{0}
  \definecolor{WHITE}{gray}{1}
  \definecolor{RED}{rgb}{1,0,0}
  \definecolor{GREEN}{rgb}{0,1,0}
  \definecolor{BLUE}{rgb}{0,0,1}
  \definecolor{CYAN}{cmyk}{1,0,0,0}
  \definecolor{MAGENTA}{cmyk}{0,1,0,0}
  \definecolor{YELLOW}{cmyk}{0,0,1,0}
 }
 \@ifundefined{definecolor}{
 }{}
 \@ifundefined{definecolor}{color}{}

\usepackage{color}

\newcommand{\be}{\begin{equation}}
\newcommand{\ee}{\end{equation}}
\newcommand{\bea}{\begin{eqnarray}}
\newcommand{\eea}{\end{eqnarray}}
\newcommand{\bse}{\begin{subequations}}
\newcommand{\ese}{\end{subequations}}

\usepackage{color}
\definecolor{d_red}{cmyk}{0.00, 0.81, 1.00, 0.27}
\definecolor{d_orange}{cmyk}{0.00, 0.33, 1.00, 0.00}
\definecolor{d_blue}{cmyk}{0.78, 0.47, 0.00, 0.20}
\definecolor{d_lgreen}{cmyk}{0.07, 0.00, 0.79, 0.29}
\definecolor{d_green}{cmyk}{0.66, 0.00, 0.71, 0.56}
\definecolor{d_blue}{cmyk}{0.78, 0.47, 0.00, 0.20}
\definecolor{d_dblue}{cmyk}{0.91, 0.79, 0.00, 0.22}
\definecolor{d_pink}{cmyk}{0.0, 0.79, 0.37, 0.29}
\definecolor{d_purple}{cmyk}{0.16, 0.54, 0.00, 0.70}
\definecolor{d_paleblue}{cmyk}{0.669, 0.338, 0.00, 0.373}
\definecolor{d_dpaleblue}{cmyk}{0.441, 0.290, 0.00, 0.580}
\definecolor{d_brown}{cmyk}{0.0, 0.490, 0.930, 0.350}
\definecolor{d_turquoise}{cmyk}{0.630, 0.04, 0.0, 0.440}
\definecolor{KIT-green}{RGB}{0, 150,130}
\definecolor{KIT-blue}{RGB}{70,100,170}

\def\bmx{\begin{pmatrix}}
\def\emx{\end{pmatrix}}

\usepackage[figure,table]{hypcap}

\makeatother

\begin{document}
\title{Breakdown of the Wiedemann-Franz law at the Lifshitz point of strained
Sr$_{2}$RuO$_{4}$ }
\author{Veronika C. Stangier}
\affiliation{Institute for Theory of Condensed Matter, Karlsruhe Institute of Technology,
Karlsruhe 76131, Germany}
\author{Erez Berg }
\affiliation{Department of Condensed Matter Physics, Weizmann Institute of Science,
Rehovot, 76100, Israel}
\author{J\"org Schmalian}
\affiliation{Institute for Theory of Condensed Matter, Karlsruhe Institute of Technology,
Karlsruhe 76131, Germany}
\affiliation{Institute for Quantum Materials and Technologies, Karlsruhe Institute
of Technology, Karlsruhe 76021, Germany}
\begin{abstract}
Strain tuning Sr$_{2}$RuO$_{4}$ through the Lifshitz point, where
the Van Hove singularity of the electronic spectrum crosses the Fermi
energy, is expected to cause a change in the temperature dependence of the electrical
resistivity from its Fermi liquid behavior $\rho\sim T^{2}$ to $\rho\sim T^{2}{\rm log}\left(1/T\right)$, a behavior consistent with experiments by Barber {\em et al.} [Phys. Rev. Lett. {\bf 120}, 076601 (2018)].
This expectation originates from the same multi-band scattering processes with
large momentum transfer that were recently shown to account for the
linear in $T$ resistivity of the strange metal Sr$_{3}$Ru$_{2}$O$_{7}$.
In contrast, the thermal resistivity $\rho_{Q}\equiv T/\kappa$, where
$\kappa$ is the thermal conductivity, is governed by qualitatively distinct
processes that involve a broad continuum of compressive modes, i.e. long wavelength density
excitations in Van Hove systems. While these compressive modes do not affect the charge
current, they couple to thermal transport and yield $\rho_{Q}\propto T^{3/2}$.
As a result, we predict that the Wiedemann-Franz law in strained Sr$_{2}$RuO$_{4}$
should be violated with a Lorenz ratio $L\propto T^{1/2}{\rm log}\left(1/T\right)$.
We expect this effect to be observable in the temperature and strain regime where
the anomalous charge transport was established.
\end{abstract}
\maketitle

\section{Introduction}

Sr$_{2}$RuO$_{4}$ is a fascinating material that combines electronic
correlations and unconventional superconductivity
whose mechanism has yet to be understood\cite{Mackenzie2003,Mackenzie2017}. Major progress in our understanding
of this material was achieved through the application of uniaxial
stress. This leads to a more than two-fold increase in the superconducting
transition temperature\cite{Hicks2014} to $T_{\rm c}^{\rm max}\approx 3.5\, {\rm K}$, a rich phase diagram where the superconducting and time reversal breaking transitions are split\cite{Grinenko2021}, and puzzling behavior
of thermodynamic properties\cite{Li2021}. The maximum in the superconducting
transition temperature occurs for strain values $\varepsilon_{xx}^{{\rm *}}\approx0.45$. 
This is  at or very near the Lifshitz transition\cite{Lifshitz1960} where 
a Van Hove singularity of the electronic spectrum crosses
the Fermi energy. 

Evidence that the normal state of the system is equally affected by
this Lifshitz point was given by Barber \emph{et. al}\cite{Barber2018}
in measurements of the electrical resistivity as function of strain
$\varepsilon_{xx}$. The main finding of these measurements are as
follows: For strain values below and above $\varepsilon_{xx}^{{\rm *}}$,
the resistivity shows Fermi liquid behavior with $\rho\approx\rho_{0}+AT^{2}$,
while for $\varepsilon_{xx}\approx\varepsilon_{xx}^{*}$ the resistivity
is more singular. The data  between $T_{c}$ and about $40\:{\rm K}$ are consistent 
with  $\rho=\rho_{0}+A T^{2}{\rm log}\left(T_{0}/T\right)$.
Over most of this temperature regime, the residual resistivity $\rho_{0}$ is a small fraction of the total resistivity.
 Hence,  the system can be safely analyzed in the clean limit. Similar results were obtained by tuning the system to the Van Hove singularity by La$^{3+}$ substituted ${\rm Sr}_{2-y}{\rm La}_{y}{\rm RuO}_{4}$\cite{Kikugawa2004,Shen2007} or epitaxial strain\cite{Burganov2016}, however with somewhat larger values for the residual resistivity $\rho_{0}$.
 
 In this paper we determine the electrical and thermal transport behavior due to electron-electron scattering  of clean Sr$_{2}$RuO$_{4}$ near the strain-induced Lifshitz point. Distinct scattering processes, both impacted by the presence of a Van Hove singularity at the Fermi energy, affect charge and heat transport differently, leading to a violation of the Wiedemann Franz law\cite{Wiedemann1853}. Charge transport is determined by  large momentum transfer scattering that  couples non-Van Hove to  Van Hove states. Heat transport is dominated by  long-wavelength scattering due to compressive modes that build a broad continuum due to the saddle point in the energy dispersion.  For perfectly clean systems and ignoring the onset of superconductivity or other ordered states  this  behavior  should continue down to lowest temperatures.  The Wiedemann Franz law should only be recovered once impurity scattering becomes dominant.
 As the two  scattering rates that govern electrical and thermal transport are both caused by electron-electron interactions and the presence of the Van Hove singularity, it seems natural that the  violation of the Wiedemann Franz law occurs in the same temperature regime  $3.5 \:{\rm K}\cdots 20-40\:{\rm K}$ where the deviation from the $T^2$ behavior in the resistivity was observed\cite{Barber2018}.  Our results  call for thermal conductivity measurements under strain to verify the importance of electron-electron interactions of the quasi two-dimensional sheet of the Fermi surface.

\begin{figure*}
    \centering
    \includegraphics[width=0.85\textwidth]{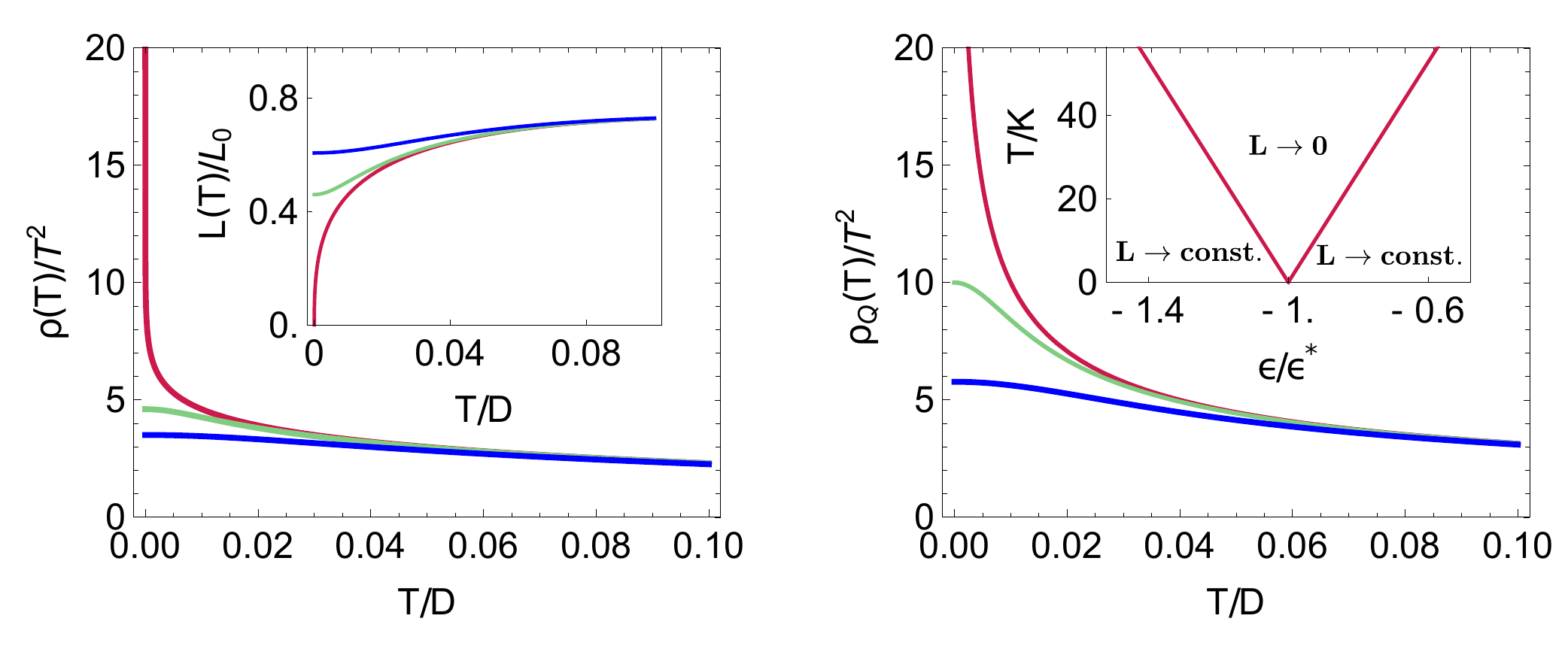}
    \caption{Temperature dependence of the resistivity $\rho(T)\sim T^2\log(D/T)$ (left) and thermal resistivity $\rho_Q(T)\sim T^{3/2}$ (right)   divided by the Fermi liquid behavior  $T^2$ at the Van Hove point (red curves). The more singular behavior of the thermal resistivity is clearly visible. $D$ is the effective bandwidth of the problem. As one moves away from the Van Hove point, one expects  a temperature scale $T^\ast$  below which both transport coefficients recover ordinary Fermi liquid behavior following Eqs\eqref{rho_strain} and \eqref{rhoQ_strain}. We use $T^\ast/D=0.01$ (green curves) and $T^\ast/D=0.03$ (blue curves). The inset shows the corresponding behavior for the Lorentz ratio $L(T)$. The right index shows the characteristic temperature scale $T^*$ for the crossover to Van Hove dominated scattering as the strain is varied. We used the parameters given in Appendix~\ref{App:tightbinding}. In this Inset we also indicate the regimes where the Lorenz number $L$ tends to a finite value or vanishes as $T$ decreases.}
    \label{fig:WF}
\end{figure*}

The existence of logarithmic corrections to the $T^{2}$ behavior
near van Hove singularities was discussed in the past with $\rho\sim T^{2}{\rm log^{2}}\left(1/T\right)$
in Ref.\cite{Hlubina1995} and $\rho\sim T^{2}{\rm log}\left(1/T\right)$
in Ref.\cite{Hlubina1996}. Including impurity scattering then changes the behavior to $\rho=\rho_0+ B T^{3/2}$, where the $T$-dependent term is, however, a small correction to the residual resistivity.
A careful analysis of the transport processes in systems with impurity scattering was recently performed in Ref.\cite{Herman2019}. Interestingly, in this work a drop in the Lorentz ratio near the Van Hove singularity was found.
These results are somewhat surprising as
usually a single, \emph{hot} point on the Fermi surface does not influence
the transport properties of a system. A prime example is the transport
behavior near a density-wave instability where hot spots are Fermi
surface points connected by the ordering vector of the density wave\cite{Hlubina1995,Stojkovic1996}.
While the scattering rate at these isolated points on the Fermi surface
is singular, the transport is dominated by generic, \emph{cold }regions of the Fermi surface that
short circuit the contribution from the hot spots and lead to Fermi liquid behavior with
$\rho\sim T^{2}$\cite{Hlubina1995}. Only the inclusion of impurity
scattering changes the behavior to $\rho=\rho_{0}+AT^{d/2}$\cite{Rosch1999,Syzranov2012}.
However, in this case the $T$-dependent term is a small correction
to the dominant residual resistivity $\rho_{0}$, in distinction to
the experimental result of Ref.\cite{Barber2018}. 

An interesting exception to
the rule that hot spots are irrelevant for transport was recently
presented in Ref.\cite{Mousatov2020} to explain the linear in
$T$ resistivity of the strange metal Sr$_{3}$Ru$_{2}$O$_{7}$.
It was shown that scattering processes $cc\leftrightarrow ch$, 
in which a cold electron ($c$) becomes hot ($h$)  after colliding with another
cold electron, exists everywhere on the Fermi surface, i.e. it cannot
be short circuited. For  Sr$_{3}$Ru$_{2}$O$_{7}$ the hot electrons
are made up of an exceptionally sharp peak in the density of states
that crosses the Fermi surface. The relevance of this scenario for
Sr$_{2}$RuO$_{4}$ was already mentioned in Ref.\cite{Mousatov2020}.
Below we show that for the specific Fermi surface geometry of Sr$_{2}$RuO$_{4}$
the  $cc\leftrightarrow ch$  processes do indeed yield the $\rho\sim T^{2}{\rm log}\left(1/T\right)$
of Ref.\cite{Hlubina1996}. One has to be careful however to include
all bands that cross the Fermi surface, as mere umklapp processes
of a single band do not provide the necessary phase space for $cc\leftrightarrow ch$
processes down to lowest temperatures. Hence, our analysis shows that
the electrical resistivity of strained Sr$_{2}$RuO$_{4}$ can be
understood in terms of the $cc\leftrightarrow ch$ approach
of Ref.\cite{Mousatov2020}.
 
The modifications of the electrical resistivity at the strain-tuned
Lifshitz point are argued to be  due to the divergent density of states at the Van Hove
singularity with large momentum transfer in the involved scattering processes. 
Sr$_2$RuO$_4$ is of course a three-dimensional material. Hence, the logarithmic divergence in the density of states is cut off at some energy scale $t_{\perp}$ set by the inter-layer hopping. However, this energy scale was shown in experiments to be only a few Kelvin\cite{Mackenzie2003}. This is consistent with  the three-dimensional electronic structure where the dispersion in the c-direction is particularly weak for in-plane momenta near the van Hove point\cite{Haverkort2008,Roising2019}. With $t_{\perp}$ comparable to $T_c$ we will ignore these effects in what follows.
In addition to these density of states effects,  electrons near a Van
Hove singularity are also expected to yield singular negative
corrections to the compressibility or other elastic constants:
\begin{equation}
\delta C\left(T\right)\sim-\frac{D}{v_{0}}\log\frac{D}{k_{B}T},\label{eq:compress}
\end{equation}
where $v_{0}$ is the unit cell volume and $D$ the
band width. Unless preempted by other states of order, such as superconductivity, this
should eventually give rise to a lattice  instability at some low temperature. 
 These compressive modes are related to a broad continuum
in the long wave-length density fluctuation spectrum of systems with Van Hove singularity\cite{Gopalan1992}.
Such fluctuations are known to give rise to singular single particle
scattering rates, with frequency and temperature dependencies that
depend on the details of the band dispersion\cite{Gopalan1992,Pattnaik1992}.
The corresponding contribution to the resistivity is small, however, since scattering of electrons from these fluctuations involves a small momentum transfer.

The presence of distinct scattering processes that do and do not contribute
to the resistivity suggests to analyze different transport properties.
After all, the Wiedemann-Franz law, according
to which the Lorenz ratio 
\begin{equation}
L=\frac{\kappa}{\sigma T}=\frac{\rho}{\rho_{Q}}
\label{eq:LLL}
\end{equation}
of the thermal conductivity $\kappa=T/\rho_{Q}$ and the electronic
conductivity $\sigma=1/\rho$ approaches $L_{0}=\frac{\pi^{2}}{3}\left(\frac{k_{B}}{e}\right)^{2}$
, is caused by the same scattering processes contributing to thermal
and charge transport\cite{Ziman1960}. This is certainly the case
for disordered electrons\cite{Castellani1987,Michaeli2009}. In clean
Fermi liquids, charge current relaxation requires umklapp scattering, while heat current relaxation
transport does not. Hence, there is no reason to expect that $L\rightarrow L_{0}$.
However, given that both scattering rates are proportional to $T^{2}$
with $\tau_{J,Q}^{-1}=A_{J,Q}T^{2}$ for charge ($J$) and heat ($Q$) transport processes, one still expects a constant
Lorenz ratio $L\left(T\rightarrow0\right)\rightarrow L_{0}\frac{A_{J}}{A_{Q}}$. 

We show that the thermal transport at the Lifshitz point is governed
by a scattering rate $\tau_{Q}^{-1}\propto T^{3/2}$ caused by the
continuum of density fluctuations, while the resistivity follows the
discussed $\tau_{J}^{-1}\propto T^{2}\log\frac{D}{T}$. As a result,
we obtain for the Lorenz ratio 
\begin{equation}
L\propto T^{1/2}\log\frac{D}{T},
\label{eq:WFV}
\end{equation}
which vanishes as $T\rightarrow0$. Hence, we expect a strong breakdown of
the Wiedemann-Franz law at the Lifshitz point of strained Sr$_{2}$RuO$_{4}$. 
These results are valid right at the Lifshitz transition. As the chemical potential moves away from the Van Hove point one expects to recover the usual Fermi liquid behavior. This is sketched in Fig.~\ref{fig:WF}, where we use 
\begin{equation}
\rho\left(T\right)\approx A_{J}T^{2}\log\frac{D}{\sqrt{T^{2}+T^{*2}}}
\label{rho_strain}
\end{equation}  and 
\begin{equation}
\rho_{Q}\left(T\right)\approx A_{Q}T^{2}\left(\frac{D^{2}}{T^{2}+T^{*2}}\right)^{1/4}.   
\label{rhoQ_strain}
\end{equation}
 The energy scale $k_{\rm B}T^{*}$ is essentially the distance of the Van Hove point to the Fermi energy and is shown in the right inset of Fig.~\ref{fig:WF} using realistic parameters for the electronic structure of Sr$_2$RuO$_4$; see Appendix~\ref{App:tightbinding} for details. The left inset shows the Lorentz number as function of temperature. Our prediction for the violation of the Wiedemann-Franz law is consistent with the numerical solution of the Boltzmann equation of Ref.\cite{Herman2019}, where a suppression in the Lorentz number for weakly disordered systems was seen near the Van Hove point.
 
 It is of interest to contrast the behavior found here with the one of two-dimensional Fermi liquids without Van Hove singularity. Then the single-particle scattering rate is enhanced by a logarithmic term, compared to the usual $T^2$ behavior $\tau^{-1}_{\rm qp}\sim T^2\log\left(D/T\right)$. As this enhancement is due to small momentum transfer processes, it will not affect the resistivity\cite{Pal2012}, i.e. $\rho(T)\sim T^2$. However, the thermal conductivity does couple to  forward scattering processes and acquires an additional logarithmic contribution  $\rho_Q(T)\sim T^2\log\left(D/T\right)$\cite{Lyakhov2003}. Hence one also finds a violation of the Wiedemann-Franz law  $L(T)\sim1/\log\left(D/T\right)$. This violation, however, is much weaker than the one we predict at a Van Hove point.

In what follows we briefly discuss the electronic structure of Sr$_{2}$RuO$_{4}$
within a three-band model. We then summarize the behavior of long
wavelength density fluctuations near a Van Hove point. Finally we
present our results for the charge and heat transport. In the appendices we comment briefly on the relation to Matthiessen's rule,
summarize the tight-binding parametrization of the band structure, determine the single particle scattering rate, and present
our results for the current relaxation rate $\tau_{J}^{-1}\left(\omega\right)$ in the regime $\tau_{J}^{-1}(\omega=0)\ll \omega \ll D$.

\section{Model and Density response}

As we include multi-band effects in our analysis we start from the
following three-band model with kinetic energy 
\begin{equation}
H_{0}=\sum_{\boldsymbol{k}\sigma}\psi_{\boldsymbol{k}\sigma}^{\dagger}{\cal H}\left(\boldsymbol{k}\right)\psi_{\boldsymbol{k}\sigma}
\end{equation}
where $\psi_{\boldsymbol{k}\sigma}=\left(d_{\boldsymbol{k},xy,\sigma},d_{\boldsymbol{k},xz,\sigma},d_{\boldsymbol{k},yz,\sigma}\right)^{T}$
with annihilation operators for electrons in the Ru $4d_{xy}$ as
well as $4d_{xz}$ and $4d_{yz}$ orbitals, respectively. For our analysis we use the single particle Hamiltonian 
\begin{equation}
{\cal H}\left(\boldsymbol{k}\right)=\left(\begin{array}{ccc}
\varepsilon_{\boldsymbol{k}xy} & 0 & 0\\
0 & \varepsilon_{\boldsymbol{k}xz} & V_{\boldsymbol{k}}\\
0 & V_{\boldsymbol{k}} & \varepsilon_{\boldsymbol{k}yz}
\end{array}\right),
\end{equation}
where we employ the dispersion relations obtained in Ref.\cite{Burganov2016}
from angular-resolved photoemission data for the unstrained system
and follow Ref.\cite{Barber2018} to account for the changes in the
dispersion at finite applied strain. 
Uniaxial strain $\epsilon_{xx}$ lifts the degeneracy between states at momenta $(\pi,0)$ and $(0,\pi)$ and splits the Van Hove singularity into two peaks. For $\epsilon_{xx}^\ast=0.45\%$ the Van Hove singularity at 
\begin{equation}
\boldsymbol{k}_{\rm VH}=(0,\pi)
\end{equation} 
crosses the Fermi energy.
The details of this analysis
are summarized in Appendix~\ref{App:tightbinding}. In Fig.\ref{fig:densityOfStates} we show our results for the
strain dependence of the Fermi surface and the density of states. 
\begin{figure}
    \centering
    \includegraphics[width=0.45\textwidth]{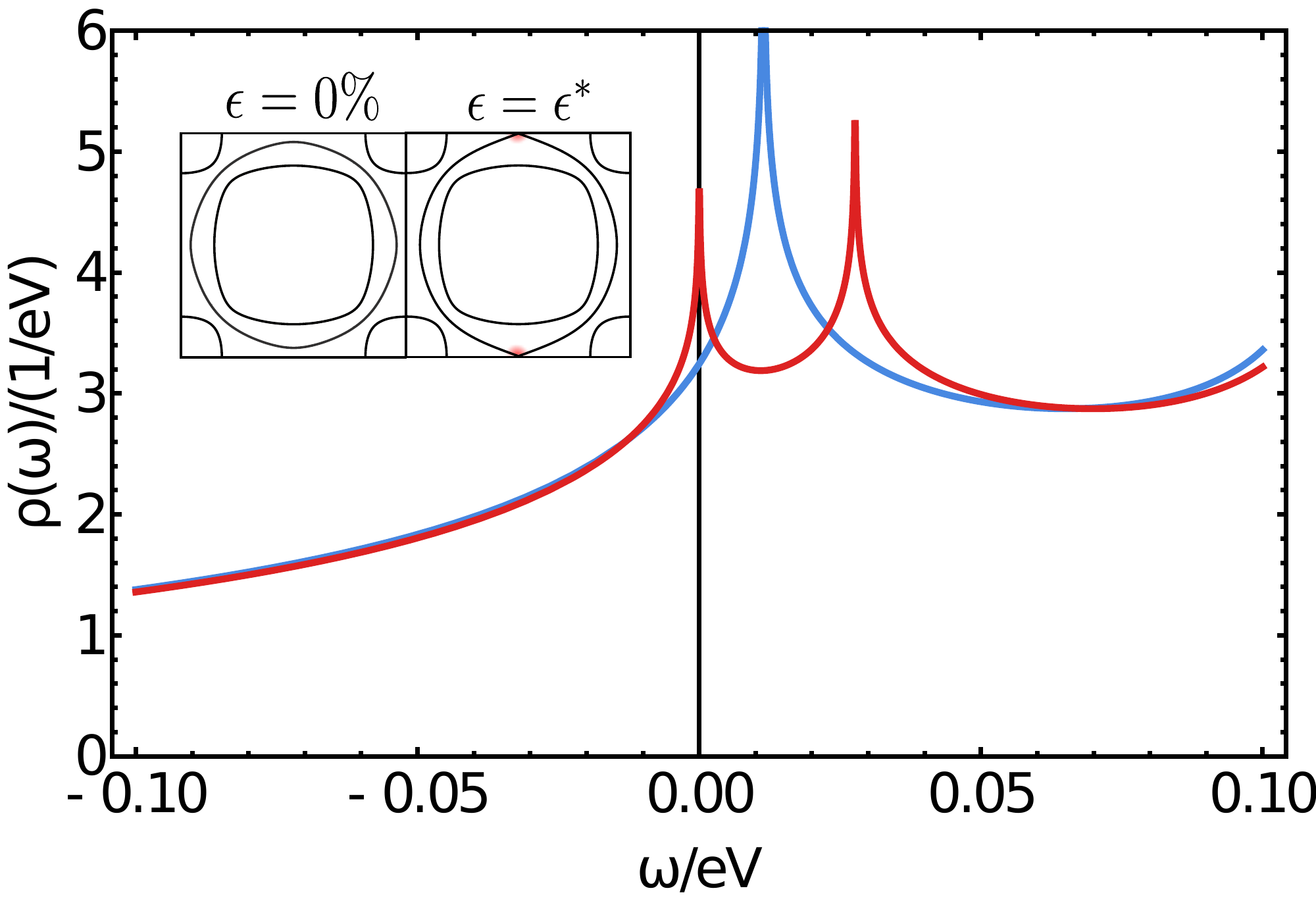}
    \caption{Density of states of the $\gamma$ band at zero strain (blue) and at the strain value $\epsilon^\ast$ that corresponds to the Lifshitz point where one of the strain-split Van Hove singularities crosses the Fermi energy (red). The inset shows the Fermi energies at zero strain (left) and at the Lifshitz point (right), where the hot parts of the Fermi surface are indicated in red.}
    \label{fig:densityOfStates}
\end{figure}

\begin{figure}
    \centering
    \includegraphics[width=0.45\textwidth]{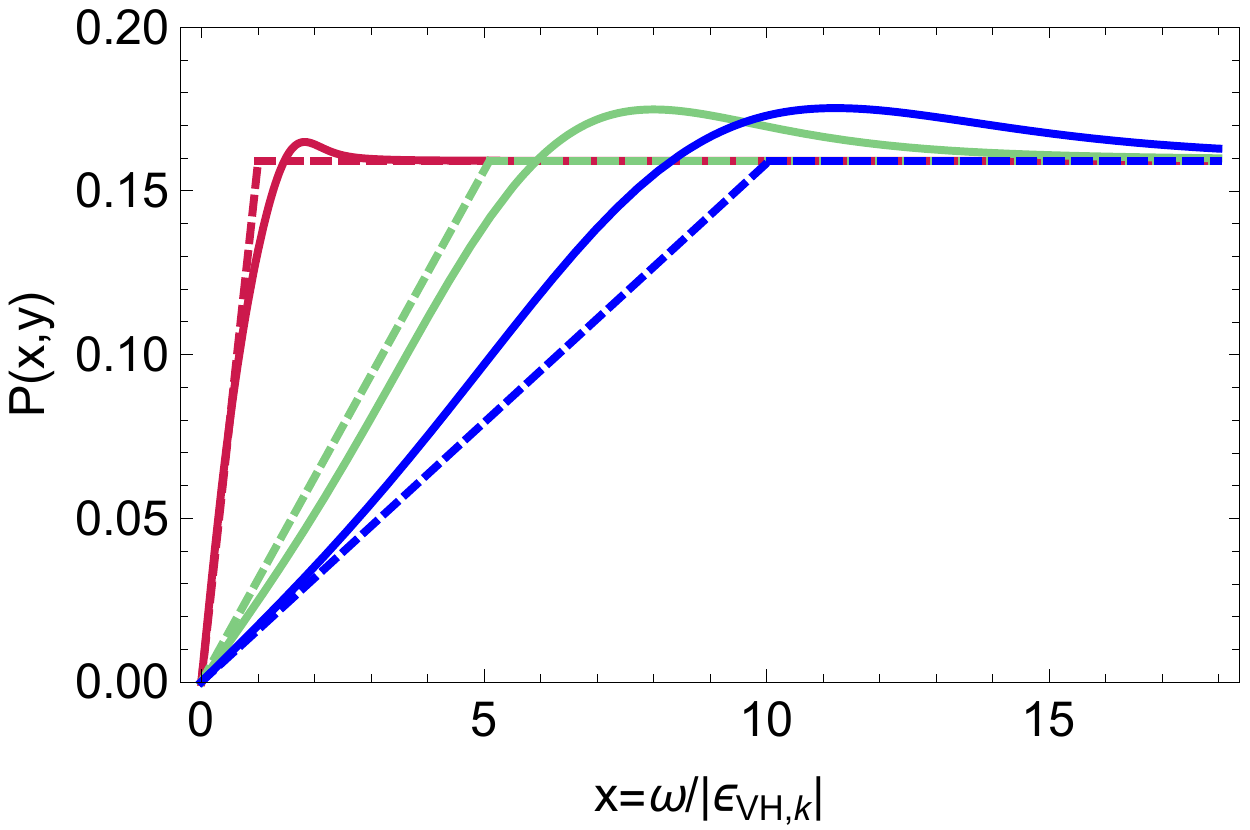}
    \caption{Scaling function $P(x,y)$ of Eq.~\eqref{eq:density responseT2} (solid lines) determining the frequency and temperature dependence of the charge excitation spectrum near a Van Hove point as function of $x=\omega/\left|\varepsilon_{{\rm VH},\boldsymbol{q}}\right|$, in comparison with the simplified version of Eq.~\eqref{eq:density response} together with the substitution  Eq.~\eqref{thermaleffect} (dashed lines) to include thermal effects. We used $m=1$ and show results for $y=0.1$ (red) $y=5$ (green), and $y=10$ (blue), where $y=T/\left|\varepsilon_{{\rm VH},\boldsymbol{q}}\right|$.}
    \label{fig:ScalFunct}
\end{figure}

The details of this electronic structure will be important when we
analyze the kinematics of umklapp scattering events that are crucial
for the electrical resistivity. For the thermal transport long-wavelength
density excitations of a system with Van Hove singularity will become
important. The density excitation spectrum follows from
\begin{eqnarray}
{\rm Im}\Pi\left(\boldsymbol{q},\omega\right) & = & \int\frac{d^{2}p}{4\pi}\left(f\left(\varepsilon_{\boldsymbol{p}+\boldsymbol{q}}\right)-f\left(\varepsilon_{\boldsymbol{p}}\right)\right)\nonumber \\
 & \times & \delta\left(\omega-\varepsilon_{\boldsymbol{p}+\boldsymbol{q}}+\varepsilon_{\boldsymbol{p}}\right).
 \label{eq:bubble}
\end{eqnarray}
As the low momentum regime is dominated by states near the saddle
point of the dispersion, we approximate the so called $\gamma$ band
with dispersion $\varepsilon_{\boldsymbol{k}xy}$ by
\begin{equation}
\varepsilon_{\boldsymbol{k}_{\rm VH}+\boldsymbol{p},xy}\approx \varepsilon_{{\rm VH},{\boldsymbol{p}}}=\frac{p_{x}^{2}-p_{y}^{2}}{2m}.
\label{eq:VH_disp}
\end{equation}
At $T=0$ the momentum integration can be performed easily, yielding\cite{Gopalan1992}
\begin{equation}
{\rm Im}\Pi\left(\boldsymbol{q},\omega\right)=-\frac{m}{2\pi}\left\{ \begin{array}{ccc}
\frac{\omega}{\left|\varepsilon_{{\rm VH},\boldsymbol{q}}\right|} & {\rm if} & \left|\omega\right|<\left|\varepsilon_{{\rm VH},{\boldsymbol{q}}}\right|\\
{\rm sign}\left(\omega\right) & {\rm if} & \left|\omega\right|>\left|\varepsilon_{{\rm VH},{\boldsymbol{q}}}\right|
\end{array}\right. .
\label{eq:density response}
\end{equation}
In comparison to the usual electronic spectrum, where the density
response vanishes for $\left|\omega\right|>v_{F}q$, a flat continuum
extends up to an energy scale of the order of the electronic bandwidth
$D$. At finite temperatures follows 
\begin{equation}
{\rm Im}\Pi\left(\boldsymbol{q},\omega\right)=P\left(\frac{\omega}{\left|\varepsilon_{{\rm VH},\boldsymbol{q}}\right|},\frac{T}{\left|\varepsilon_{{\rm VH},\boldsymbol{q}}\right|}\right),
\label{eq:density responseT1}
\end{equation}
with scaling function
\begin{eqnarray}
P\left(x,y\right) & = & \sqrt{y}\frac{m}{4\sqrt{\pi}}\left(Li_{1/2}\left(-e^{\frac{1}{4y}\left(2x-1\right)^{2}+\frac{2x}{y}}\right)\right.\nonumber \\
 & - & \left.Li_{1/2}\left(-e^{\frac{1}{4y}\left(2x-1\right)^{2}}\right)\right)
 \label{eq:density responseT2}
\end{eqnarray}
and polylogarithmic function $Li_{s}\left(z\right)$. As we show in
Fig.~\ref{fig:ScalFunct}, the finite temperature density response can be expressed to a good approximation
in the form Eq.\eqref{eq:density response} but with 
\begin{equation}
\left|\varepsilon_{{\rm VH},\boldsymbol{q}}\right|\rightarrow\Omega_{\boldsymbol{q}}\left(T\right)=\sqrt{\varepsilon_{{\rm VH},\boldsymbol{q}}^{2}+\left(k_{B}T\right)^{2}}.
\label{thermaleffect}
\end{equation}
The corresponding real part of the density response yields $\Pi\left(\boldsymbol{q},0\right)=-\frac{m}{\pi^{2}}\log\frac{D}{\Omega_{\boldsymbol{q}}\left(T\right)}$
which leads to Eq.~\eqref{eq:compress} for the compressibility.

An important result that follows from this continuum of density excitations
is an anomalous single particle scattering rate\cite{Gopalan1992,Pattnaik1992}.
To see this we analyze the imaginary part of the single-particle self
energy coupled to the above density fluctuations:
\begin{eqnarray}
{\rm Im}\Sigma\left(\boldsymbol{k},\omega\right) & = & 2\frac{U^{2}}{N}\sum_{\boldsymbol{k}'}\left(f_{0}\left(\varepsilon\left(\boldsymbol{k}'\right)\right)+n_{0}\left(\varepsilon\left(\boldsymbol{k}'\right)-\omega\right)\right)\nonumber \\
 & \times & {\rm Im}\Pi\left(\boldsymbol{k}-\boldsymbol{k}',\omega-\varepsilon\left(\boldsymbol{k}'\right)\right),
 \label{eq:self energy}
\end{eqnarray}
where $U$ is the electron-electron interaction and $f_0$ and $n_0$ are the Fermi and Bose distribution functions, respectively.
While the momentum transfer $\boldsymbol{k}-\boldsymbol{k}'$ is small,
the individual Fermi momenta $\boldsymbol{k}$ and $\boldsymbol{k}'$
do not necessarily have to be located in the vicinity of the Van Hove
point. Indeed, the result ${\rm Im}\Sigma\left(\boldsymbol{k},\omega\right)\propto\left|\omega\right|^{\gamma}$
for the single particle self energy depends sensitively whether $\boldsymbol{k}$
and $\boldsymbol{k}'$ are near or away from the saddle point of the
dispersion. In the former case the dispersion $\varepsilon\left(\boldsymbol{k}\right)$
is given by Eq.\eqref{eq:VH_disp}. In this case follows $\gamma=1$
which is the result obtained in Ref.\cite{Pattnaik1992}.
Alternatively
we can analyze the self energy for generic momenta. Now it is sufficient
to assume a parabolic spectrum for $\varepsilon\left(\boldsymbol{k}\right)$,
which yields $\gamma=3/2$\cite{Gopalan1992}. A third option for
momenta with parabolic dispersion and Fermi velocity parallel to the
directions of zeros of Eq.\eqref{eq:VH_disp} yields a somewhat more
singular behavior with $\gamma=4/3$. We will show that the behavior
with $\gamma=3/2$ is the one that determines the thermal conductivity.
Further details of the analysis of the single particle self energy are summarized in Appendix~\ref{App:selfenergy}.

\section{Electrical Resistivity}
To calculate the resistivity we use the standard Boltzmann ansatz
\begin{equation}
   \frac{\partial f_{\boldsymbol{k}}}{\partial t}+e\boldsymbol{E}\cdot\frac{\partial f_{\boldsymbol{k}}}{\partial\boldsymbol{k}}=-{\cal C}_{\boldsymbol{k}}\left[f\right]
   \end{equation}
with scattering operator ${\cal C}_{\boldsymbol{k}}\left[f\right]$ and determine the charge current 
\begin{equation}
\boldsymbol{j}=-\frac{e}{N}\sum_{\boldsymbol{k}\sigma}\boldsymbol{v}_{\boldsymbol{k}}f_{\boldsymbol{k}}
\end{equation}
at given electrical field.  Next we expand $f_{\boldsymbol{k}} $ for small deviations from equilibrium, parametrized by a function $\psi_{\bm{k}}$ (proportional to the electric field $\bm{E}$):
\begin{equation}
f_{\boldsymbol{k}}=f_{0}\left(\varepsilon_{\boldsymbol{k}}\right)-T\frac{\partial f_{0}\left(\varepsilon_{\boldsymbol{k}}\right)}{\partial\varepsilon_{\boldsymbol{k}}}\psi_{\boldsymbol{k}}.
\end{equation}
The linearized collision operator due to electron-electron scattering up to second order in $U$ takes the usual form
\begin{eqnarray} 
{\cal C}_{\boldsymbol{k}_{1}}\left[\psi\right]	 &=&	\frac{2\pi}{\hbar}U^{2}\sum_{\boldsymbol{k}_{2}\boldsymbol{k}_{3}\boldsymbol{k}_{4}}\delta\left(\varepsilon_{\boldsymbol{k}_{1}}+\varepsilon_{\boldsymbol{k}_{2}}-\varepsilon_{\boldsymbol{k}_{3}}-\varepsilon_{\boldsymbol{k}_{4}}\right) \nonumber \\
&\times & L_{\varepsilon_{\boldsymbol{k}_{1}},\varepsilon_{\boldsymbol{k}_{2}},\varepsilon_{\boldsymbol{k}_{3}},\varepsilon_{\boldsymbol{k}_{4}}} \sum_{\boldsymbol{G}}\delta_{\boldsymbol{k}_{1}+\boldsymbol{k}_{2}-\boldsymbol{k}_{3}-\boldsymbol{k}_{4}-\boldsymbol{G}}
\nonumber \\
	&\times&	\left(\psi_{\boldsymbol{k}_{1}}+\psi_{\boldsymbol{k}_{2}}-\psi_{\boldsymbol{k}_{3}}-\psi_{\boldsymbol{k}_{4}}\right).
	\label{eq:collission}
\end{eqnarray}
The sum over $\boldsymbol{G}$ goes over the entire reciprocal lattice. In practice only a few terms contribute since four vectors of the first Brilouin zone only reach a finite number of reciprocal lattice vectors. For notational simplicity we suppressed the band index in the collision integral and assumed the same interaction matrix element $U$ for intra- and inter-band scattering. Below, we will discuss in detail which bands are included in the analysis. The phase space restrictions of degenerate fermions are included through
\begin{equation}
L_{\varepsilon_{1},\varepsilon_{2},\varepsilon_{3},\varepsilon_{4}}=f_{0}\left(-\varepsilon_{1}\right)f_{0}\left(-\varepsilon_{2}\right)f_{0}\left(\varepsilon_{3}\right)f_{0}\left(\varepsilon_{4}\right).
\end{equation}
Obviously $\psi_{\boldsymbol{k}}\propto \rm{const.} $ is a zero mode of the collision operator due to charge conservation. Similarly, $\psi_{\boldsymbol{k}}\propto\varepsilon_{\boldsymbol{k}} $ is a zero mode due to energy conservation. However, $\psi_{\boldsymbol{k}}\propto k_{\alpha}$ is not a zero mode as the momenta do not have to add up to zero. Umklapp scattering processes spoil momentum conservation and yield a finite resistivity.

When analyzing the resistivity we make the simplifying assumption that 
\begin{equation}
\psi_{\boldsymbol{k}}\approx \psi_{0}e\boldsymbol{E}\cdot\boldsymbol{v}_{\boldsymbol{k}}
\label{eq:approx_psi}
\end{equation}
is determined by the velocity $\boldsymbol{v}_{\boldsymbol{k}}=\frac{\partial\varepsilon_{\boldsymbol{k}}}{\partial\boldsymbol{k}}$.  This assumption is only justified if a given scattering process is kinematically allowed everywhere on the Fermi surface. Otherwise the coefficient $\psi_{0}$ will depend strongly on the position on the Fermi surface, effects that occur as  corrections of  the current vertex in a diagrammatic treatment of transport.  As a result the longest scattering rate that is allowed on the entire Fermi surface  dominates the transport behavior; it short circuits stronger scattering events that are kinematically  only allowed on a subset of the Fermi surface. While this {\em hot spot} reasoning is well established\cite{Hlubina1995,Hlubina1996,Stojkovic1996,Rosch1999,Syzranov2012,Mousatov2020}, it appears at odds with the expectation based on Matthiessen's rule~\cite{Matthiessen1864} where one has to add scattering rates, not times; see Appendix~\ref{App:Resistivity}.
 With the ansatz Eq.~\eqref{eq:approx_psi} we finally arrive at the following result for the low-frequency ( $\omega < \tau^{-1}_J$) conductivity
 \begin{equation}
 \sigma \left(\omega \right)=\frac{n_D}{-i\omega+\tau_J^{-1}},
 \label{eq:Drude}
 \end{equation}
with Drude weight $n_D=\frac{e^{2}}{N}\sum_{\boldsymbol{k}\sigma}\frac{\partial^{2}\varepsilon_{\boldsymbol{k}}}{\partial k_{\alpha}\partial k_{\beta}}f_{0}\left(\varepsilon_{\boldsymbol{k}}\right)$ and scattering rate for charge transport 
\begin{eqnarray}
\tau_{J}^{-1}&=&\frac{2\pi}{\hbar}\frac{U^{2}e^{2}}{4n_{D}k_{{\rm B}}T}\sum_{\boldsymbol{k}_{1}\cdots\boldsymbol{k}_{4}}\delta\left(\varepsilon_{\boldsymbol{k}_{1}}+\varepsilon_{\boldsymbol{k}_{2}}-\varepsilon_{\boldsymbol{k}_{3}}-\varepsilon_{\boldsymbol{k}_{4}}\right)\nonumber \\
&\times & L_{\varepsilon_{\boldsymbol{k}_{1}},\varepsilon_{\boldsymbol{k}_{2}},\varepsilon_{\boldsymbol{k}_{3}},\varepsilon_{\boldsymbol{k}_{4}}} \sum_{\boldsymbol{G}}\delta_{\boldsymbol{k}_{1}+\boldsymbol{k}_{2}-\boldsymbol{k}_{3}-\boldsymbol{k}_{4}-\boldsymbol{G}}
\nonumber \\
&\times & \left(v_{\boldsymbol{k}_{1}}+v_{\boldsymbol{k}_{2}}-v_{\boldsymbol{k}_{3}}-v_{\boldsymbol{k}_{4}}\right)^{2}.
\end{eqnarray}
Due to the velocity term, momentum conserving scattering processes vanish in the single-band case and, as expected,  do not contribute to the resistivity. For multiple bands momentum-conserving terms can, of course, change the current.

We perform the momentum summation according to 
\begin{equation}
\frac{1}{N}\sum_{\boldsymbol{k}}F\left(\boldsymbol{k}\right)  \approx\sum_{S}\int d\epsilon\rho_{S}\left(\epsilon\right)\int_{S}\frac{d\varphi}{2\pi}F\left(\epsilon,\varphi\right).
\end{equation}
In the last step we sum over segments of the Fermi surface with essentially constant density of state $\rho_{S}\left(\epsilon\right)\approx \rho_F$ or with logarithmic density of state $\rho_{S}\left(\epsilon\right)\approx \rho_{F}\log\left(D/|\epsilon|\right)$. 
Finally we use 
\begin{eqnarray}
I&=&\int d\epsilon_{1}\cdots d\epsilon_{4}\delta\left(\epsilon_{1}+\epsilon_{2}-\epsilon_{3}-\epsilon_{4}\right) L_{\epsilon_{1},\epsilon_{2},\epsilon_{3},\epsilon_{4}} \nonumber \\
&\times & \prod_{i=1}^{4}\rho_{S_{i}}\left(\epsilon_{i}\right)
\approx \frac{2\pi^{3}}{3}\left(k_{B}T\right)^{3}\rho_{F}^4 \log^{n}\frac{D}{T},
\end{eqnarray}
where $n=0\cdots 4$ is the number of electrons near a Van Hove point that are involved in a given scattering process. For example, for a $cc\rightarrow ch$ process, $n=1$. 

Depending on the number of such hot electrons involved,  distinct scattering processes contribute differently to the resistivity. At first glance, the contribution of a scattering process with $n$ hot electrons seems to be $\rho^{(n)}\propto T^2\log^n\left(1/T\right)$, which would be dominated by $\rho^{(4)}$ at low $T$, i.e. by the largest value of $n$. However, as discussed above, we must  i) ensure the existence of umklapp scattering for given $n$, ii) check that those processes exist on the entire Fermi surface, and iii) ensure that some of the states involved have a finite velocity, needed to carry the current.  We will see that this yields $n=0$ for a single band problem, while $n=1$ if one includes additional bands  that  cross the Fermi energy. Hence, including multi-band scattering events, the final result for the electrical resistivity will be 
\begin{equation}
    \rho(T)= A_J T^2\log\left(D/T\right),
    \label{eq:resistivity}
\end{equation}
with $A_J\sim U^2 \rho_F^3/n_D$. 
Eq.\eqref{eq:Drude} also includes the low frequency Drude response of the system where  $\tau_J$  corresponds to the zero frequency limit of the charge transport rate. In Appendix~\ref{App:memory} we also discuss   $\tau_J^{-1}(\omega) $ in the opposite limit  $\tau_J^{-1}(0)\ll \omega \ll D$ within a memory function approach and obtain 
\begin{equation}
\tau_J^{-1}(\omega)\sim \omega^2 \log\left(D/|\omega|\right),
\end{equation}
which determines the optical conductivity at these intermediate frequencies
\begin{equation}
    \sigma\left(\omega\right)=\frac{n_{D}}{-i\omega\left(1+\lambda\left(\omega\right)\right)+\tau_{J}^{-1}\left(\omega\right)}.
    \label{sigma_memory}
\end{equation}
Here, $\lambda\left(\omega\right)=\lambda_{0}\left(\log2-\frac{\omega}{D}\arctan\left(\frac{D}{\omega}\right)\right)$ is the optical mass enhancement that follows from Kramers-Kronig transformation. This frequency dependency parallels the temperature dependency of the optical scattering rate at low frequencies.

For a problem with two Van Hove points  at $\boldsymbol{k}^{(1)}_{\rm VH}=(\pi,0)$ 
and $\boldsymbol{k}^{(2)}_{\rm VH}=(0,\pi)$, i.e.  without the problems that occur in the single-band model relevant to Sr$_2$CuO$_4$ Eq.\eqref{eq:resistivity} was already obtained in Ref.\cite{Hlubina1996}. In what follows we discuss the kinematics of the single-band and multi-band problems separately.

\begin{figure*}
    \centering
    \includegraphics[width=0.85\textwidth]{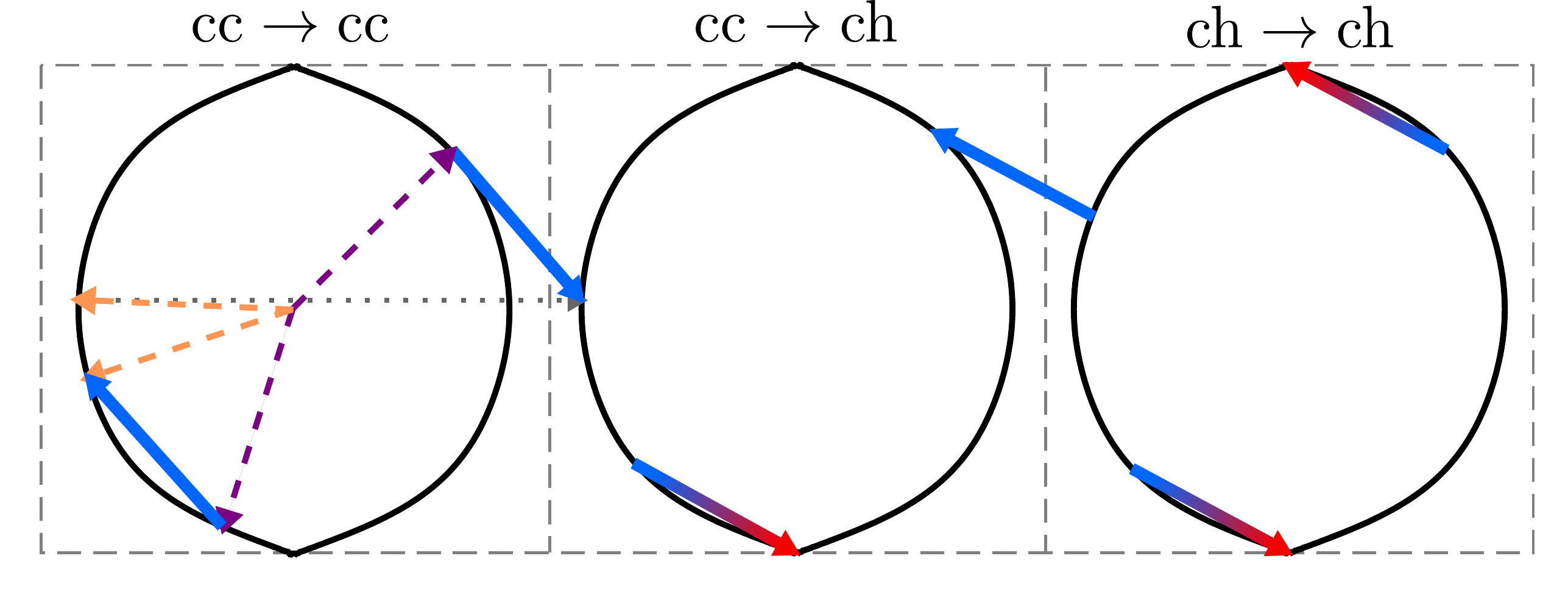}
    \caption{A $\mathrm{cc\rightarrow cc}$ process, where two electrons scatter from two cold initial states (purple) into two cold final states (orange) with umklapp scattering, leads to a resistivity of $\rho\propto T^2$. The blue arrows indicate the momentum transfer for each electron. The $\mathrm{cc\rightarrow ch}$ process contributes by $\rho\propto T^2\log\left(1/T\right)$ and dominates, if every point on the Fermi surface participates in this scattering. The $\mathrm{ch\rightarrow ch}$ processes always conserve momentum, therefore they do not contribute to the resistivity.}
    \label{fig:ScatProc}
\end{figure*}

\subsection{Resistivity for the $\gamma$-band only}
As we have seen, electron-electron scatterings yield different contributions to the resistivity, depending on the number of hot electrons involved. Which of these scattering processes are possible  depends on the geometry of the Fermi surface. 
For simplicity, we start from a simplified model that includes only the $\gamma$ band. In principle, the Fermi surface allows scattering processes with $n=0,1,2$ and $4$ hot electrons. Of these, only $cc\leftrightarrow cc$  (i.e. $n=0$) and $cc\leftrightarrow ch$  ($n=1$) contribute to the resistivity as they allow umklapp processes. Fig. \ref{fig:ScatProc} shows the possible scatterings schematically for zero, one and two hot electrons.

The $cc\leftrightarrow ch$ scattering dominates the resistivity if every point on the cold Fermi surface can participate in such a scattering process. These processes have to be umklapp processes in order to contribute to the resistivity. It turns out that this is not possible for the entire Fermi surface. For an electron close to the Van Hove point the momentum transfer is not sufficiently large to surpass the gap between two neighboring Fermi surfaces. For the tight-binding parameters relevant for Sr$_2$RuO$_4$, we obtain that more than $10\%$ of the 
 Fermi surface cannot participate in $cc\leftrightarrow ch$ scattering processes.  This implies that our assumption of Eq.~\eqref{eq:approx_psi} for the distribution function is not justified. 
 
States on the Fermi surface   relatively close to the Van Hove point cannot participate in $cc\rightarrow ch$ umklapp scattering and should at sufficiently low temperatures 
lead to a resistivity that behaves as $\rho_{cc\rightarrow cc}\propto T^2$.  One can qualitatively understand the crossover at higher temperature by considering two competing contributions to the resistivity with 
 $\rho_{cc\rightarrow cc}$ and $\rho_{cc\rightarrow ch}\propto T^2\log\left(1/T\right)$ that add up to the total resistivity 
\begin{equation}
    \rho_{ee}^{-1}\sim \rho_{cc\rightarrow cc}^{-1}+\rho_{cc\rightarrow ch}^{-1}.
    \label{inparallel}
\end{equation}
At lowest $T$ the resistivity exhibits the usual relation $\rho\propto T^2$, while at higher temperatures the $cc\rightarrow ch$ process dominates and the resistivity obtains the logarithmic correction $\rho\propto T^2\log\left(1/T\right)$. The value of the crossover temperature can then be estimates as $T^\ast\sim D \exp\left(-c \frac{1-x}{x}\right)$ where $x$ is the fraction of the Fermi surface where $\mathrm{cc\leftrightarrow ch}$ scattering processes are kinematically forbidden and $c$ is a numerical coefficient of order unity.

\subsection{Resistivity with inter-band scattering}
\begin{figure}
    \centering
    \includegraphics[width=0.3\textwidth]{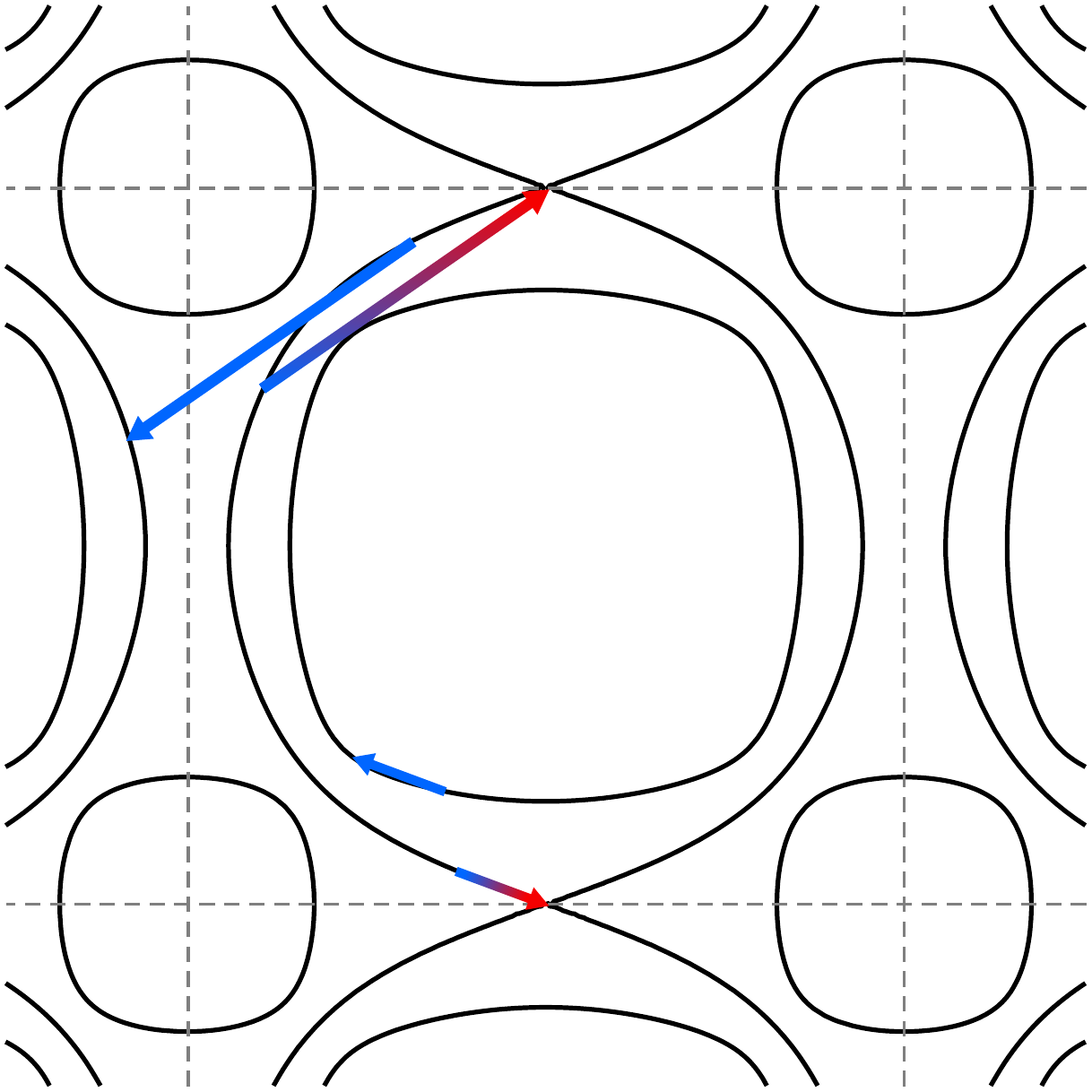}
    \caption{When all bands are considered, every point on the cold Fermi surface can participate in $\mathrm{cc\rightarrow ch}$ scattering, either by umklapp or inter band scattering.}
    \label{fig:multiBand}
\end{figure}

So far we have restricted our analysis to the $\gamma$ band. However, to explain the transport properties of $\mathrm{Sr_2RuO_4}$, one has to take the whole Fermi surface into account.  Since current is not conserved in inter-band scattering, no matter if it is umklapp or not, these scattering events contribute to the resistivity.
Since the $\beta$ band is convex and only includes cold regions (called $\mathrm{c_\beta}$), we can always find a $\mathrm{c_\beta\rightarrow c_\beta}$ process, as long as the momentum transfer is not too large. These processes fill the kinematic gap of the single-band case such that now the entire $\gamma$-band can participate in $\mathrm{c_\beta c_\gamma \rightarrow c_\beta h_\gamma}$ scattering, either by umklapp or inter band scattering, as shown in Fig. \ref{fig:multiBand}. A similar argumentation can be made for the $\alpha$ band. Hence, additional {\em cold} states on the Fermi surface that can couple via inter-band collisions open up the phase space for singular scattering processes. This yields the somewhat surprising result that more cold states suppress the ability of these states to short circuit the singular transport processes that involve Van Hove points. 
Since no new forbidden areas in the $\alpha$ and $\beta$ bands emerge,  the resistivity follows  Eq.\eqref{eq:resistivity} down to lowest temperatures.  

\section{Thermal transport}
Next, we analyze the thermal transport for a system at the Lifshitz point.
For the thermal transport we start again from the Boltzmann equation
\begin{equation}
\frac{\partial f_{\boldsymbol{k}}}{\partial t}+e\boldsymbol{v}_{\boldsymbol{k}}\cdot\frac{\partial f_{\boldsymbol{k}}}{\partial\boldsymbol{r}}=-{\cal C}_{\boldsymbol{k}}\left[f\right]
\end{equation}
with the same collision operator given in Eq.\eqref{eq:collission} for the electrical resistivity.
The heat current
\begin{equation}
\boldsymbol{j}_{Q}=\frac{1}{N}\sum_{\boldsymbol{k}\sigma}\left(\varepsilon_{\boldsymbol{k}}-\mu\right)\boldsymbol{v}_{\boldsymbol{k}}f_{\boldsymbol{k}}
\end{equation}
is then determined as function of the temperature gradient, which enters the Boltzmann equation through 
\begin{equation}
\frac{\partial f_{\boldsymbol{k}}}{\partial\boldsymbol{r}}\approx-\frac{\partial f_{0}\left(\varepsilon_{\boldsymbol{k}}\right)}{\partial\varepsilon_{\boldsymbol{k}}}\,\frac{(\varepsilon_{\boldsymbol{k}}-\mu)}{T}\frac{\partial T}{\partial\boldsymbol{r}}\,.
\end{equation}

 The crucial difference is of course that the momentum and the thermal current do not couple, i.e. there is no need to analyze whether umklapp scattering processes exist\cite{Lawrence1973,Schulz1995,Hartnoll2013,Principi2015,Hartnoll2016,comment_thermal}. Hence, small momentum transfer collisions, where $\boldsymbol{q}=\boldsymbol{k}_{3}-\boldsymbol{k}_{1}=\boldsymbol{k}_{2}-\boldsymbol{k}_{4}$ is small, contribute to the thermal resistivity. Of course, the two incoming momenta $\boldsymbol{k}_{1}$ and $\boldsymbol{k}_{2}$ can be far from each other, an issue that will be crucial when we determine the correct temperature dependence of the thermal conductivity.

We hence ignore umklapp processes in the  thermal transport and perform one momentum sum using momentum conservation for the reciprocal lattice vector $\boldsymbol{G}=\boldsymbol{0}$. This yields for time-independent thermal gradients
\begin{eqnarray}
\boldsymbol{v}_{\boldsymbol{k}_{1}}\cdot \frac{\partial f_{\boldsymbol{k}_1}}{\partial\boldsymbol{r}}&=&\frac{2\pi U^{2}}{\hbar}\sum_{\boldsymbol{k}_{2}\boldsymbol{q}}L_{\varepsilon_{\boldsymbol{k}_{1}},\varepsilon_{\boldsymbol{k}_{2}},\varepsilon_{\boldsymbol{k}_{1}+\boldsymbol{q}},\varepsilon_{\boldsymbol{k}_{2}-\boldsymbol{q}}}  \nonumber \\
&\times&\delta\left(\varepsilon_{\boldsymbol{k}_{1}}+\varepsilon_{\boldsymbol{k}_{2}}-\varepsilon_{\boldsymbol{k}_{1}+\boldsymbol{q}}-\varepsilon_{\boldsymbol{k}_{2}-\boldsymbol{q}}\right)\nonumber \\
&\times&\left(\psi_{\boldsymbol{k}_{1}}+\psi_{\boldsymbol{k}_{2}}-\psi_{\boldsymbol{k}_{1}+\boldsymbol{q}}-\psi_{\boldsymbol{k}_{2}-\boldsymbol{q}}\right)\, .
\label{eq:Bethermal}
\end{eqnarray}
We consider a generic momentum $\boldsymbol{k}_{1}$ on the Fermi surface. For small momentum transfer 
$\boldsymbol{k}_{1}+\boldsymbol{q}$   will then  be near   $\boldsymbol{k}_{1}$.  For the other incoming momentum 
$\boldsymbol{k}_{2}$ we  can, however,  assume that it is located in the vicinity of the Van Hove point, which implies that $\boldsymbol{k}_{2}-\boldsymbol{q}$ is also near that point. Following our analysis for the charge transport, one might be tempted to conclude that the resistivity has $n=2$ states near the Van Hove point, which would imply a scattering rate that behaves as $T^2\log^2\left(1/T\right)$. However,  as we saw in the analysis of the density fluctuation spectrum in  Eqs.\eqref{eq:density response} - \eqref{eq:density responseT2}, small momentum transfer processes have a much increased phase space which gives rise to a more singular behavior for the thermal transport. 

To proceed we first make the ansatz for the distribution function
\begin{equation}
\psi_{\boldsymbol{k}}=\psi_{0}\frac{\varepsilon_{k}-\mu}{T}\boldsymbol{v}_{\boldsymbol{k}}\cdot\nabla T
\label{eq:psithermal}
\end{equation}
which is analog to Eq.\eqref{eq:approx_psi} for thermal transport. As before, we have to demonstrate that a given scattering process is present everywhere on the Fermi surface in order to justify the assumption that the coefficient $\psi_{0}$ is weakly dependent on  momentum. 
With Eq.~\eqref{eq:psithermal}  and given the vanishing  velocity at the Van Hove point,  we can safely neglect $\psi_{\boldsymbol{k}_{2,4}}$ in Eq.~\eqref{eq:Bethermal}.  If we use the identity $f_{0}\left(\varepsilon\right)f_{0}\left(-\varepsilon'\right)=\left(f_{0}\left(\varepsilon\right)-f_{0}\left(\varepsilon'\right)\right)n_0\left(\omega\right)$ for $\varepsilon'=\varepsilon+\omega$  we can express the sum over  $\boldsymbol{k}_2$, which runs over states near the Van Hove point, in terms of the density spectrum of Eq.~\eqref{eq:bubble}
\begin{eqnarray}
\boldsymbol{v}_{\boldsymbol{k}_{1}}\cdot \frac{\partial f_{\boldsymbol{k}_1}}{\partial\boldsymbol{r}}&=&-\frac{2\pi U^{2}T}{\hbar N}\frac{\partial f_{0}\left(\varepsilon_{\boldsymbol{k}_{1}}\right)}{\partial\varepsilon_{\boldsymbol{k}_{1}}}
\sum_{\boldsymbol{q}}\left(\psi_{\boldsymbol{k}_{1}}-\psi_{\boldsymbol{k}_{1}-\boldsymbol{q}}\right)\nonumber \\
&\times & \left(n_0\left(\varepsilon_{\boldsymbol{k}_{1}-\boldsymbol{q}}-\varepsilon_{\boldsymbol{k}_{1}}\right)+f_0\left(\varepsilon_{\boldsymbol{k}_{1}-\boldsymbol{q}}\right)\right)\nonumber \\
&\times&{\rm Im}\Pi\left(\boldsymbol{q},\varepsilon_{\boldsymbol{k}_{1}}-\varepsilon_{\boldsymbol{k}_{1}-\boldsymbol{q}}\right).
\end{eqnarray}
Hence, we explicitly see how the coupling to compressive fluctuations affects the thermal conductivity. One now finds that, up to numerical factors of order unity, the collision operator is merely determined by the single-particle self energy of Eq.~\eqref{eq:self energy}.
\begin{equation}
\boldsymbol{v}_{\boldsymbol{k}_{1}}\cdot \frac{\partial f_{\boldsymbol{k}_1}}{\partial\boldsymbol{r}}=\tau_{\boldsymbol{k}_{1}Q}^{-1}T\frac{\partial f_{0}\left(\varepsilon_{\boldsymbol{k}_{1}}\right)}{\partial\varepsilon_{\boldsymbol{k}_{1}}}\psi_{\boldsymbol{k}_{1}}
\end{equation}
with the heat transport scattering rate
\begin{figure}
    \centering
    \includegraphics[width=0.3\textwidth]{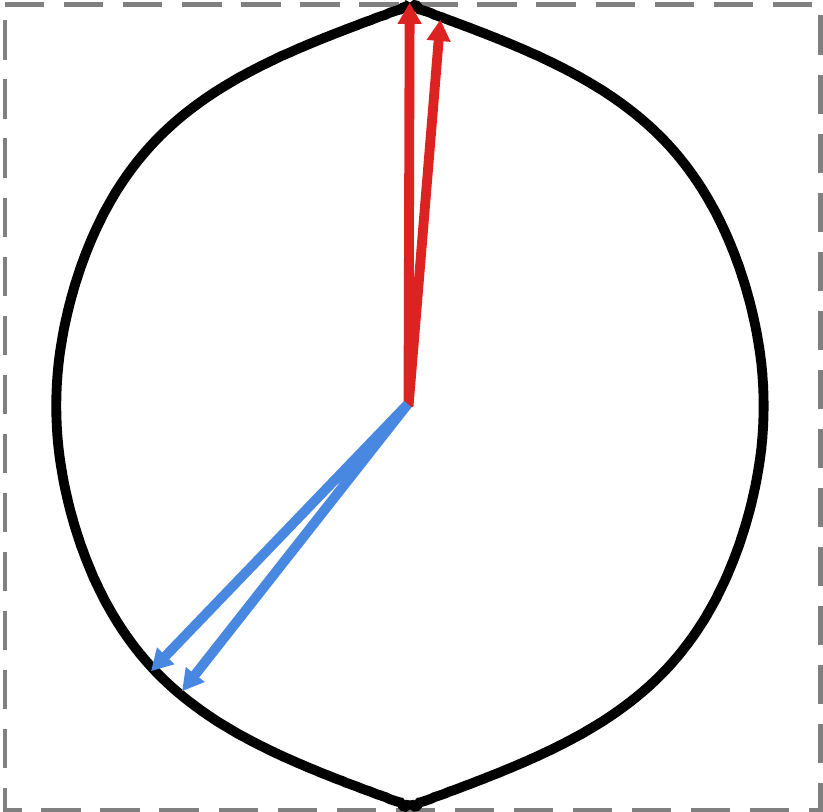}
    \caption{ch$\rightarrow$ch process dominating the thermal transport with two hot momenta near the Van Hove point (red) and two cold electrons with generic dispersion (blue). For small momentum transfer, this process is allowed everywhere on the Fermi surface.}
    \label{fig:chch}
\end{figure}
\begin{figure}
    \centering
    \includegraphics[width=0.45\textwidth]{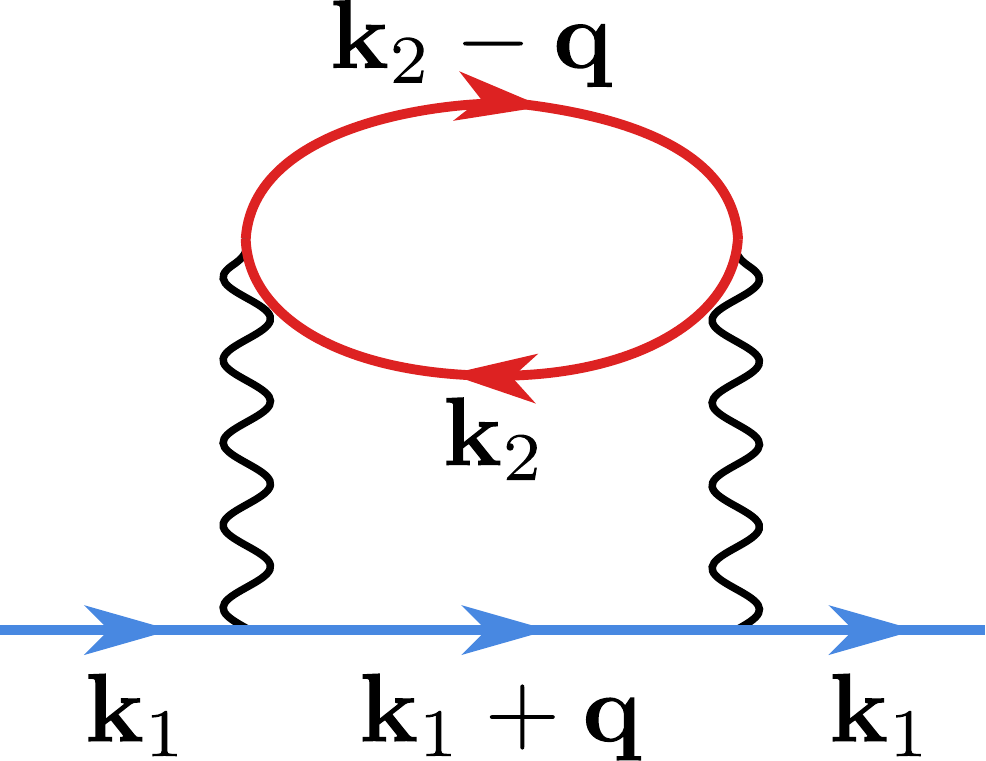}
    \caption{Self energy diagram determining the thermal resistivity with hot momenta $\mathbf{k}_2$ and $\mathbf{k}_2-\mathbf{q}$ with hyperbolic dispersion (red) and cold momenta $\mathbf{k}_1$ and $\mathbf{k}_1+\mathbf{q}$ with parabolic dispersion (blue).}
    \label{fig:Feynman}
\end{figure}
\begin{equation}
\tau_{\boldsymbol{k}_{1} Q}^{-1}\approx - 2\mathrm{Im}\Sigma\left(\boldsymbol{k}_{1}, \varepsilon_{_{\boldsymbol{k}_{1}}}\right).
\label{eq:ThResistivity}
\end{equation}
This process is kinematically allowed everywhere on the Fermi surface, and amounts to an analysis of the self energy diagram shown in Fig. \ref{fig:Feynman}. In Appendix~\ref{App:selfenergy} we summarize the analysis of the self energy, see also Refs.\cite{Gopalan1992,Pattnaik1992}. While there exists more singular scattering in some parts of the Fermi surface, a generic point obeys
\begin{equation}
\tau_{\boldsymbol{k},Q}^{-1}=\frac{16U^{2}\rho_{F}^2}{3\sqrt{D}}T^{3/2}.
\end{equation}
This  yields for the thermal conductivity 
\begin{equation}
\kappa=\frac{1}{TN}\sum_{\boldsymbol{k}\sigma}v_{\boldsymbol{k}}^{2}\tau_{\boldsymbol{k},Q}\left(-\frac{\partial f_{0}\left(\varepsilon_{\boldsymbol{k}}\right)}{\partial\varepsilon_{\boldsymbol{k}}}\right)\left(\varepsilon_{\boldsymbol{k}}-\mu\right)^{2}.
\end{equation}
Performing the integrals in the usual manner yields for the thermal resistivity introduced in Eq.~\eqref{eq:LLL} the result
\begin{equation}
\rho_Q=A_Q D^{1/2} T^{3/2},
\end{equation}
where the coefficient $A_Q$ is of the same order of magnitude as $A_J$ that determines  the charge transport in Eq.~\eqref{eq:resistivity}.
This finally yields the $T$-dependence of the Lorentz number given in Eq.~\eqref{eq:WFV} and the concomitant violation of the Wiedemann Franz law.

\section{Conclusions}
In conclusion, we analyzed the electrical and thermal transport of
clean Sr$_{2}$RuO$_{4}$ under strain near the Lifshitz point where
a Van Hove singularity of the $\gamma$-sheet of the Fermi surface
crosses the Fermi energy. Based on the observation of well defined
quasiparticles in this material we perform a quasi-classical Boltzmann
transport theory. For larger frequencies, we also present results
for the optical conductivity within a memory function approach. We
find that both, the electrical and the thermal transport are affected
by the vicinity to the Lifshitz point. The known result for the electrical
resistivity, discussed already in Ref.\cite{Hlubina1996}, is physically interpreted
in terms of scattering processes where an electron near the Van Hove
point collides with a \emph{cold} electron away from the Van Hove
point and scatters into two other cold states. This gives rise to
a logarithmic enhancement of the charge transport rate $\tau_{J}^{-1}\sim T^{2}\log\left(1/T\right)$
and hence of the electrical resistivity, consistent with observations
of Refs.\cite{Kikugawa2004,Shen2007,Burganov2016,Barber2018}. In particular, the observation by Barber et al.\cite{Barber2018}, with
a very small residual resistivity $\rho_{0}$, demonstrate that an
understanding of these results has to be achieved without resorting
to impurity scattering processes that usually increase the phase space
for singular electron-electron scattering\cite{Rosch1999}. For the specific
electronic structure of Sr$_{2}$RuO$_{4}$ we show that the logarithmic
enhancement is present down to lowest temperatures if one includes
inter-band scattering processes. The reason is that intra-band scattering
requires umklapp processes which have more stringent phase space
requirements. Hence, the anomalous transport behavior seen in Sr$_{2}$RuO$_{4}$
is particularly robust as there are several bands that cross the Fermi
surface. The situation is drastically different if one analyses thermal
transport. Systems with Van Hove singularities are characterized by
a much enhanced phase space of long-wavelength compressive modes.
These compressive modes do not couple to charge transport. However,
they are able to relax the thermal current and give rise to a thermal
transport rate $\tau_{Q}^{-1}\propto T^{3/2}$. As a consequence the
Wiedemann-Franz law is violated with a Lorentz number that vanishes
with a powerlaw plus logarithmic corrections. The experimental confirmation
of this prediction would in particular demonstrate the importance
of the mentioned compressive modes for the low-energy excitations
of Sr$_{2}$RuO$_{4}$. This would also be of interest as it would
be curious to study whether these long-wavelength modes might enhance
an existing or even cause an instability towards a superconducting
state. 
A related issue is clearly the role of higher-order processes not included in our analysis. The fact that the compressibility is guaranteed to vanish at a finite temperature is clear evidence that the system will eventually undergo some kind of instability. Our transport theory is however expected to be valid in the regime that leads up to this instability. 

Our prediction for a violation of the Wiedemann Franz law is only
valid as long as one is in the clean regime. At some low temperature,
impurity scattering effects should become important. Then we expect
to recover at lowest temperatures the Wiedemann-Franz law where $L$
approaches $L_{0}=\frac{\pi^{2}}{3}\left(k_{B}/e\right)^{2}$. For
the leading low-temperature corrections, still dominated by impurity
scattering events, one expects $\tau_{Q,J}^{-1}=\tau_{{\rm imp}}^{-1}+C_{Q,J}T^{3/2}$
i.e. both transport rates should follow the same temperature dependence.
This behavior could be relevant for the observations of La-substituted
systems\cite{Shen2007,Burganov2016}. Again, it's confirmation would give strong evidence for the
importance of compressive modes for the low-energy electronic degrees
of freedom.
\begin{acknowledgments}
We are grateful to M. Garst, S. A. Hartnoll, C. Hicks, and A. P. MacKenzie for useful discussions.  V. C. S. and J. S. were supported by the Deutsche Forschungsgemeinschaft
(DFG, German Research Foundation) - TRR 288-422213477 Elasto-Q-Mat
(project B01). E. B. was supported by the
European Research Council (ERC) under grant HQMAT (Grant Agreement No. 817799), the Minerva foundation, and a Research grant from Irving and Cherna Moskowitz.
\end{acknowledgments}

\appendix

\section{Matthiesen's rule and hot versus cold carriers}
\label{App:Resistivity}
According to Matthiesen\textquoteright s rule, the total resistivity
is the sum of resistivities of different scattering mechanisms, $\rho_{{\rm tot}}=\rho_{{\rm imp}}+\rho_{{\rm e-ph}}+\rho_{{\rm e-e}}$,
meaning that the dominant contribution to the resistivity comes from
the shortest life time. This seems in sharp contrast to what was  discussed
in Eq.\eqref{inparallel} and to our strategy to ensure that a given scattering mode
is not short circuited by    scattering processes on the Fermi surface with smaller rate.
To clarify this issue we consider the linearized Boltzmann equation
in the operator form 
\begin{equation}
\hat{C}\left|\psi\right\rangle =\left|S\right\rangle ,
\end{equation}
where $\hat{C}$ is the collision operator, $\psi_{\boldsymbol{k}}=\left\langle \boldsymbol{k}\mid\psi\right\rangle $
the correction to the distribution function and $S_{\boldsymbol{k}}=\left\langle \boldsymbol{k}\mid S\right\rangle $
some external source term. The source term due to an external electric
field is $S_{\boldsymbol{k}}\equiv-\frac{e}{T}\boldsymbol{E}\left(\boldsymbol{q},\omega\right)\cdot\boldsymbol{v}_{\boldsymbol{k}}$.
With the collision integral ${\cal C}_{\boldsymbol{k}}$ that we
used in the main text, the operator is defined as
\begin{equation}
\hat{C}\psi_{\boldsymbol{k}}=\frac{1}{-T\frac{\partial f^{\left(0\right)}\left(\varepsilon_{\boldsymbol{k}}\right)}{\partial\varepsilon_{\boldsymbol{k}}}}\int_{\boldsymbol{k}'}\frac{\delta{\cal C}_{\boldsymbol{k}}}{\delta\psi_{\boldsymbol{k}'}}\psi_{\boldsymbol{k}'}.
\end{equation}
It is a Hermitian operator with respect to the inner product
\begin{equation}
\left\langle \phi\mid\psi\right\rangle =\int_{\boldsymbol{k}}w_{\boldsymbol{k}}\phi_{\boldsymbol{k}}^{*}\psi_{\boldsymbol{k}}
\end{equation}
with weight function $w$$_{\boldsymbol{k}}=-T\frac{\partial f^{\left(0\right)}\left(\varepsilon_{\boldsymbol{k}}\right)}{\partial\varepsilon_{\boldsymbol{k}}}>0$.
The entropy production can then we written as $\partial S/\partial t=\left\langle \psi\left|\hat{C}\right|\psi\right\rangle $,
i.e. the operator is positive definite. The eigenvalues of the collision
operator are the scattering rates 
\begin{equation}
\hat{C}=\sum_{\lambda}\left|\lambda\right\rangle \tau_{\lambda}^{-1}\left\langle \lambda\right|,
\end{equation}
where $\lambda$ labels distinct modes of the scattering process.
For a rotation-invariant problem they could for example describe different
angular momentum modes. 

If the system is governed by several, distinct scattering mechanism,
such as impurity, electron-phonon, or electron-electron scattering,
the individual operators add up 
\[
\hat{C}=\hat{C}_{{\rm imp}}+\hat{C}_{{\rm el-ph}}+\hat{C}_{{\rm el-el}}.
\]
We write this as $\tau_{\lambda}^{-1}=\sum_{s}\tau_{\lambda,s}^{-1}$
where the index $s$ indicates the scattering mechanisms. In order
to determine the distribution function we now have to invert the collision
operator in the  subspace orthogonal to zero modes that indicate
conserved quantities such as particle number or energy. This is possible
as long as these zero modes are orthogonal to the source term $\left|S\right\rangle $.
It follows
\begin{eqnarray}
\left|\psi\right\rangle  & = & \hat{C}^{-1}\left|S\right\rangle \nonumber \\
 & = & \sum_{\lambda}'\left|\lambda\right\rangle \frac{1}{\sum_{s}\tau_{\lambda,s}^{-1}}\left\langle \lambda\mid S\right\rangle ,
\end{eqnarray}
where the prime in the sum indicates that the inversion
has been performed in the subspace orthogonal to the mentioned zero modes.
In many instances, the scattering rates for non-zero-modes depend
weakly on the eigen-mode index $\lambda$. Then one obtains the total
scattering rate
\begin{equation}
\tau_{{\rm tot}}^{-1}=\sum_{s}\tau_{s}^{-1}.
\end{equation}
This leads to Matthiesen\textquoteright s rule. However, if one considers
a system with one dominant scattering mechanism, e.g. a clean system
with electron-electron scattering where phonons are frozen out at
low temperatures, as we assume in  this paper, then
\begin{equation}
\left|\psi\right\rangle =\sum_{\lambda}\left|\lambda\right\rangle \tau_{\lambda,{\rm el-el}}\left\langle \lambda\mid S\right\rangle .
\end{equation}
For systems where distinct eigenmodes of the scattering process under
consideration vary strongly, as happens for strongly momentum
dependent scattering rates, the largest scattering time dominates
and we obtain a relation like given in Eq.\eqref{inparallel}.

To summarize, an electron that is exposed to different scattering
mechanisms must cope with all of them and the rates add according
to Matthiesen\textquoteright s rule; different scattering mechanisms
act like resistors in series. On the other hand, if there are different
modes of one dominant mechanism, the longest scattering times dominate
the conductivity; different modes act like resistors that are in parallel
where the weakest scattering short circuits the transport. 

\section{Tight binding parameters and strain dependence of the electronic
structure}
\label{App:tightbinding}

We consider the electronic structure of Sr$_{2}$RuO$_{4}$ under
arbitrary in-plane strain. The Van Hove singularity near the Fermi energy is due to the  so called $\gamma$ band, made of $4d_{xy}$
states is:
\begin{eqnarray}
\varepsilon_{\boldsymbol{k},xy} & = & -2t_{1,x}\cos k_{x}-2t_{1,y}\cos k_{y}-\mu.\\
 & - & 2\left(t_{4,+}\cos\left(k_{x}+k_{y}\right)+t{}_{4,-}\cos\left(k_{x}-k_{y}\right)\right).\nonumber 
\end{eqnarray}
It is natural to assume that for strained systems $t_{1,x}$ only depends on the change
of the nearest neighbor distance $a_{x}$, $t_{1,y}$ only on $a_{y}$,
while $t_{4,+}$ and $t_{4,-}$ are functions of the diagonal distance
$d_{+}$ and counter-diagonal distance $d_{-}$, respectively. For
simplicity we only consider in-plane strain. The nearest neighbor
distances of the strained system are $a_{\alpha}\approx a_{0}\left(1+\epsilon_{\alpha\alpha}\right)$
and $d_{\pm}\approx\sqrt{2}a_{0}\left(1+\frac{1}{2}\left(\epsilon_{xx}+\epsilon_{yy}\pm2\epsilon_{xy}\right)\right).$
Following Ref.\cite{Barber2018} it is reasonable to use $t_{1,\alpha}=t_{1}\left(1-\alpha\epsilon_{\alpha\alpha}\right)$
and $t_{4,\pm}/t_{4}=1-\frac{\alpha}{2}\left(\epsilon_{xx}+\epsilon_{yy}\pm2\epsilon_{xy}\right).$
Assuming $\sigma_{xx}$ along the $x$ direction follows $\epsilon_{xy}=0$
and $\epsilon_{yy}=-\nu_{xy}\epsilon_{xx}$ with Poisson ratio $\nu_{xy}$
such that 

\begin{eqnarray}
t_{1,x} & = & t_{1}\left(1-\alpha\epsilon_{xx}\right),\nonumber \\
t_{2,y} & = & t_{2}\left(1+\alpha\nu_{xy}\epsilon_{xx}\right),\nonumber \\
t_{4,\pm} & = & t_{4}\left(1-\frac{\alpha}{2}\left(1-\nu_{xy}\right)\epsilon_{xx}\right).\label{eq:parameters-1-1}
\end{eqnarray}
Using the elastic constants of Ref.\cite{Paglione2002} yields $\nu_{xy}=0.39$.
From Ref.\cite{Burganov2016} we take $t_{1}=0.119\,{\rm eV}$ and
$t_{4}=0.41t_{1}$ as well as $\mu=1.48t_{0}$.

The $4d_{xz}$ and $4d_{yz}$ orbitals form the so-called $\alpha$
and $\beta$ sheets of the Fermi surface. They are characterized by
a dispersion 
\begin{equation}
h=\left(\begin{array}{cc}
\varepsilon_{xz} & V\\
V & \varepsilon_{yz}
\end{array}\right)
\end{equation}
with
\begin{eqnarray}
\varepsilon_{\boldsymbol{k},xz} & = & \varepsilon_{x}^{\left(0\right)}-2t_{2,x}\cos k_{x}-2t_{3,y}\cos k_{y}-\mu, \\
\varepsilon_{\boldsymbol{k},yz} & = & \varepsilon_{y}^{\left(0\right)}-2t_{2,y}\cos k_{y}-2t_{3,x}\cos k_{x}-\mu,\nonumber \\
V_{\boldsymbol{k}} & = & -2t_{5,+}\cos\left(k_{x}+k_{y}\right)+2t_{5,-}\cos\left(k_{x}-k_{y}\right) \nonumber
\end{eqnarray}
with $\text{\ensuremath{t_{2,3,\alpha}}}=t_{2,3}\left(1-\beta\epsilon_{\alpha\alpha}\right)$
and $t_{5,\pm}/t_{5}=1-\frac{\beta}{2}\left(\epsilon_{xx}+\epsilon_{yy}\pm2\epsilon_{xy}\right)$.
For the hopping elements of the unstrained system we use the values
given in Ref.\cite{Burganov2016}: $t_{2}=0.165\:{\rm eV},$ $t_{3}=0.08t_{2}$,
and $t_{5}=0.13t_{2}$. Finally, for the unstrained system holds $\varepsilon_{x}^{\left(0\right)}-\mu=\varepsilon_{y}^{\left(0\right)}-\mu=-0.18t_{2}$.

\section{Single particle self energy}
\label{App:selfenergy}

In this appendix we analyze the single particle self energy of Eq. \eqref{eq:ThResistivity}
that describes the scattering of electrons with momentum $\boldsymbol{k}_{1}$
with density fluctuations described by Eqs.~\eqref{eq:density response}-\eqref{eq:density responseT2}. We consider the imaginary
part of the self energy as given in Eq.~\eqref{eq:self energy}.

We want to analyse a generic point on the Fermi surface, away from
the Van Hove singularity. Given the small momentum transfer we assume that $\boldsymbol{k}$
and $\boldsymbol{k}'$ are points nearby on the Fermi surface of similar
length $k_{F}$. Thus, we can safely assume a constant density of
states $\rho_{F}$ for these states. As follows from Eq.\eqref{eq:density responseT1}, the momentum
dependence of the density fluctuation spectrum is determined solely
by $\varepsilon_{{\rm VH},\boldsymbol{k}-\boldsymbol{k}'}$ with saddle
point dispersion $\varepsilon_{{\rm VH},\boldsymbol{k}}$ of Eq.\eqref{eq:VH_disp}.
If we use $\boldsymbol{k}=k_{F}\left(\cos\theta,\sin\theta\right)$
and $\boldsymbol{k}'=k_{F}\left(\cos\phi,\sin\phi\right)$, we obtain
for small $\varphi=\theta-\phi$ that
\begin{equation}
\varepsilon_{{\rm VH},\boldsymbol{k}-\boldsymbol{k}'}\approx-\varepsilon_{{\rm VH},\boldsymbol{k}}\varphi^{2}+\frac{k_{x}k_{y}}{4m}\varphi^{3}.
\end{equation}

We first analyze the self energy at $T=0$ as function of frequency.
Considering without restriction $\omega>0$ the condition coming from
the Fermi and Bose functions is that $0<\epsilon_{c}\left(\boldsymbol{p}\right)<\omega.$
This yields
\begin{equation}
{\rm Im}\Sigma\left(\boldsymbol{k},\omega\right)=2U^{2}\rho_{F}\int_{0}^{\omega}d\varepsilon_{c}\int_{0}^{2\pi}d\varphi{\rm Im}\Pi\left(\boldsymbol{k}-\boldsymbol{p},\varepsilon_{c}\right).
\end{equation}
Let us first consider momenta $\boldsymbol{k}$ with $\varepsilon_{{\rm VH},\boldsymbol{k}}\neq0$,
i.e. $\theta\neq\left(2l+1\right)\frac{\pi}{4}$. With $s\equiv\left|\varepsilon_{{\rm VH},\boldsymbol{k}}\right| \varphi^{2}\approx D\varphi^2$
follows 
\begin{eqnarray}
{\rm Im}\Sigma\left(\boldsymbol{k},\omega\right) & = & -\frac{U^{2}\rho_{F}}{\sqrt{D}}\int_{0}^{\omega}d\epsilon\int_{0}^{\infty}\frac{ds}{\sqrt{s}}{\rm Im}\Pi\left(\boldsymbol{k}-\boldsymbol{p},\epsilon\right)\nonumber \\
 & = & -\frac{U^{2}\rho_{F}}{\sqrt{D}}\int_{0}^{\omega}d\epsilon\left(\int_{\epsilon}^{\infty}\frac{ds  \epsilon} {s^{3/2}}+\int_{0}^{\epsilon}\frac{ds}{s^{1/2}}\right)\nonumber \\
 & = & -\frac{8U^{2}\rho_{F}^2}{3\sqrt{D}}\left|\omega\right|^{3/2}.
\end{eqnarray}
For the special momenta with $\theta=\left(2l+1\right)\frac{\pi}{4}$,
where $\varepsilon_{{\rm VH},\boldsymbol{k}}=0$ one should use $s\propto\varphi^{3}$
and it follows ${\rm Im}\Sigma\left(\boldsymbol{k},\omega\right)\propto\left|\omega\right|^{4/3}$.
Those results were earlier obtained in Ref.\cite{Gopalan1992}. If, instead, the dispersion
$\varepsilon_{c}\left(\boldsymbol{k}\right)$ is assumed to be equal
to $\mathbf{\varepsilon_{{\rm VH},\boldsymbol{k}}}$ one finds ${\rm Im}\Sigma\left(\boldsymbol{k},\omega\right)\propto\left|\omega\right|$.
This result was obtained in Ref.\cite{Pattnaik1992}.

At finite temperatures the analysis proceeds in an analogous manner
and one obtains 
\begin{equation}
{\rm Im}\Sigma\left(\boldsymbol{k},\varepsilon_{\boldsymbol{k}}\right)\approx-\frac{8U^{2}\rho_{F}^2}{3\sqrt{D}}T^{3/2}
\label{SET}
\end{equation}
for generic momenta and corresponding results for special directions. This result is crucial for the thermal transport behavior.

\section{Charge transport rate within the memory function formalism}
\label{App:memory}
The optical conductivity $\mathbf{j}\left(\omega\right)=\sigma\left(\omega\right)\mathbf{E}\left(\omega\right)$ obtained by the Kubo formula is 
\begin{equation}
	\sigma_{\alpha\beta}\left(\omega\right)=\frac{i}{\omega}\left(\left<T_{\alpha\beta}\right>+\chi_{j_\alpha j_\beta}\left(\omega\right)\right)\label{eq:Sigma}
\end{equation}
with inverse mass tensor  $T_{\alpha\beta}=\frac{e^2}{N}\sum_{\mathbf{k}\sigma}\frac{\partial^2\varepsilon\left(\mathbf{k}\right)}{\partial k_\alpha \partial k_\beta}c_{k\sigma}^\dagger c_{k\sigma}$ and current-current susceptibility $\chi_{j_\alpha j_\beta}$. If the current is conserved this reduces to $\sigma_{\alpha\beta}\left(\omega\right)=\frac{i\left<T_{\alpha\beta}\right>}{\omega}$. Then the real part of the conductivity diverges as $\omega$ approaches zero, making it impossible to include a current-relaxing interaction by perturbation theory in the low frequency limit. 
This problem can be tackled by introducing the memory matrix\cite{Goetze72} as a correction to the $\omega^{-1}$-behavior and regain the low frequency expansion of the conductivity
\begin{equation}
\sigma_{\alpha\beta}\left(\omega\right)=\frac{i\left<T_{\alpha\beta}\right>}{\omega+M_{\alpha\beta}\left(\omega\right)}.
\end{equation}
The imaginary part of the memory function determines the scattering rate $\tau_J^{-1}\left(\omega\right)$. The memory function contains the current-relaxing processes and can more easily  be treated within perturbation theory. By comparing with Eq. \eqref{eq:Sigma} we find that the memory function up to first order in the interaction is 
\begin{equation}
M_{\alpha\beta}\left(\omega\right)\approx \frac{\omega\chi_{j_\alpha j_\beta}\left(\omega\right)}{\left<T_{\alpha\beta}\right>}.
\end{equation}
This expansion is valid for $\omega \gg \tau_J^{-1}(0)$.
We can rewrite this as
\begin{equation}
M_{\alpha\beta}\left(\omega\right)\approx \frac{\left<\left[F_\alpha F_\beta\right]\right>-\chi_{F_\alpha F_\beta}\left(\omega\right)}{\omega\left<T_{\alpha\beta}\right>},
\end{equation}
where $F_\alpha=\left[j_\alpha, H\right]$ follows from  the equation of motion for $\chi_{j_\alpha j_\beta}\left(\omega\right)$. The scattering rate is therefore given by
\begin{equation}
\tau_J^{-1}\left(\omega\right)=\frac{-\mathrm{Im}\chi_{F_\alpha F_\beta}\left(\omega\right)}{\omega\left<T_{\alpha\beta}\right>}
\end{equation}
with
\begin{eqnarray}
\chi_{F_\alpha F_\beta}\left(\omega\right)&=&e^2U^2\sum_{\mathbf{k}_1...\mathbf{k}_4}
\left(v_{k_1}+v_{k_2}-v_{k_3}-v_{k_4}\right)^2\nonumber\\
&\times&L_{\varepsilon_{\mathbf{k}_1}\varepsilon_{\mathbf{k}_2}\varepsilon_{\mathbf{k}_3}\varepsilon_{\mathbf{k}_4}}
\sum_\mathbf{G}\delta_{\mathbf{k}_1+\mathbf{k}_2-\mathbf{k}_3-\mathbf{k}_4-\mathbf{G}}\nonumber\nonumber\\
&\times & \left[\frac{1}{\omega+\varepsilon_{\mathbf{k}_1}+\varepsilon_{\mathbf{k}_2}-\varepsilon_{\mathbf{k}_3}-\varepsilon_{\mathbf{k}_4}}\nonumber\right.\\
&&-\left.\frac{1}{\omega-\varepsilon_{\mathbf{k}_1}-\varepsilon_{\mathbf{k}_2}+\varepsilon_{\mathbf{k}_3}+\varepsilon_{\mathbf{k}_4}}\right].
\end{eqnarray}
The imaginary part of the susceptibility can be evaluated in analogy to $\tau_J\left(0\right)$, where the external frequency replaces temperature. Taking into account the proper phase space for umklapp scattering,  this yields a frequency depending scattering rate 
\begin{equation}
\tau_J^{-1}\left(\omega\right)=\lambda_0\frac{\omega^2}{D}\log\left(\frac{D}{\left|\omega\right|}\right).
\label{rate_memory}
\end{equation}
This result, valid in the regime $\tau_J^{-1}(\omega=0)\ll \omega \ll D$, gives rise to the optical conductivity given in Eq.\eqref{sigma_memory}.
The rate is therefore determined by the same processes that give rise to the temperature dependence of the resistivity. As was discussed in Ref.\cite{Rosch2000,Rosch2002} for higher frequencies, shorter rates that do not rely on the condition of momentum conservation may also affect the frequency dependence of the optical conductivity. Here, the shorter single-particle rate $\sim \omega^{3/2}$ could become relevant. For our problem those would only become relevant once $\omega$ becomes comparable to the band width $D$. Thus is unclear whether such an intermediate regime will be observable in experiment. At lowest frequencies the result Eq.\eqref{rate_memory} is however the dominant one.


\begin{thebibliography}{10}
\bibitem{comment_thermal} {\color{black}Strictly speaking, the thermal resistivity of a clean system is non-zero in the absence of umklapp scattering if it is measured under conditions of zero electrical current.}
\end{thebibliography}
\end{document}